\documentclass[final,5p,times,twocolumn]{elsarticle}

\usepackage[colorlinks = true, pdfpagelabels]{hyperref}
\usepackage{amssymb}
\usepackage{subfig}
\usepackage{textcomp}
\usepackage{graphicx}
\usepackage{upgreek}
\usepackage{gensymb}
\usepackage{doi}
\usepackage[version=3]{mhchem}
\usepackage{sidecap}
\usepackage{upgreek}
\usepackage{float}
\usepackage[utf8]{inputenc}
\usepackage{booktabs}
\usepackage{lineno}
\journal{Nucl. Instr. Meth. B }
\biboptions{sort&compress}

\begin{document}

\begin{frontmatter}

\title{Characterization of $^{233}$U alpha recoil sources for $^{229(m)}$Th beam production}

\author[jyv]{I. Pohjalainen\corref{cor1}}
\ead{ilkka.pohjalainen@jyu.fi}
\author[jyv]{I.D. Moore\corref{cor1}}
\ead{iain.d.moore@jyu.fi}
\author[jyv]{T. Sajavaara}

\address[jyv]{University of Jyvaskyla, Department of Physics, P.O. Box 35 (YFL), FI-40014 University of Jyvaskyla, Finland}

\cortext[cor1]{Corresponding author.}

\begin{abstract}
Radioactive~$^{233}$U alpha recoil sources are being considered for the production of a thorium ion source to study the low-energy isomer in~$^{229}$Th with high-resolution collinear laser spectroscopy at the IGISOL facility of the University of Jyväskylä. In this work two different~$^{233}$U sources have been characterized via alpha and gamma spectroscopy of the decay radiation obtained directly from the sources and from alpha-recoils embedded in implantation foils. These measurements revealed rather low~$^{229}$Th recoil efficiencies of only a few percent. Although the low efficiency of one of the two sources can be attributed to its inherent thickness, the low recoil efficiency of the second, thinner source, was unexpected. Rutherford backscattering spectrometry (RBS) was performed to investigate the elemental composition as a function of depth revealing a contamination layer on top of the thin source. The combination of spectroscopic methods proves to be a useful approach in the assessment of alpha recoil source performance in general.
\end{abstract}

\begin{keyword}
  radioactive source \sep alpha-recoil \sep Th-229 \sep U-233

\PACS 29.25.Rm

\end{keyword}

\end{frontmatter}

\section{Introduction}

Currently, there is a significant lack of information on ground state nuclear structure properties of the isotopes in the actinide region obtained through high-resolution optical spectroscopy~\cite{Campbell2016}. Therefore, the production of actinide ion beams has become a focus of recent efforts at the IGISOL facility of the Accelerator Laboratory, University of Jyväskylä~\cite{Voss2017}. In particular, the isotope of~$^{229}$Th has been of considerable interest in recent years due to its exceptionally low-energy isomeric state of only $7.8\pm 0.5$~eV~\cite{Beck2007, Beck2009}. After the initial (indirect) detection of the isomer~\cite{Wense2016}, the electromagnetic moments and nuclear charge radii have been measured with high-resolution optical spectroscopy~\cite{Thielking2018}. There are ongoing efforts at IGISOL to create a thorium ion source to produce~$^{229}$Th ion beams in different atomic charge states in order to perform a high-resolution study with the collinear laser spectroscopy facility~\cite{Vormawah2018}.

A convenient method to create~$^{229}$Th and populate both the ground- and the low-lying isomeric state is to use~$^{233}$U, which alpha decays directly to~$^{229}$Th with a 2.1(5)\% branching ratio~\cite{Thielking2018, Barci2003} to the isomeric state. By placing a source of~$^{233}$U into a gas cell and stopping the recoils of the alpha decay in buffer gas, an off-line source for~$^{229}$Th ions can be realized with the possibility to extract the ions in several charge states. $^{233}$U has a long half-life of 1.6$\cdot 10^5$ years~\cite{Jaffey1974} and therefore a source can be manufactured and used without any expected depletion of the material.

The somewhat low branching value to the isomeric state has the unfortunate consequence that the source strength has to be sufficiently large enough to meet the required ion beam intensity for a fluorescence-based laser spectroscopy experiment. Assuming a requirement of $\sim$1000 ions of~$^{229m}$Th per second, a~$^{233}$U source strength of 50~kBq would be needed. Taking into account losses due to recoil efficiency, stopping and extraction from the gas cell may result in a source activity of several MBq.

The benefit of using an off-line alpha-recoil emitter compared to on-line production is the absence of a primary beam-induced plasma, which can be a significant source of free electrons causing recombination and neutralization of the ions of interest~\cite{Moore2010}. Although there is a significant amount of alpha radiation emanating from the~$^{233}$U sources, ionization of the buffer gas is insignificant. If a~$^{229}$Th ion is stopped in helium gas, the high ionization potential of helium (24.6 eV) ensures that a helium atom can only donate electrons to thorium if the charge state is above 3+. If the buffer gas is additionally purified from contaminant molecules which have lower ionization potentials, the~$^{229}$Th ions can remain in a 3+ charge state during extraction from a gas cell.

\begin{figure}
  \centering
  \includegraphics[width=\columnwidth]{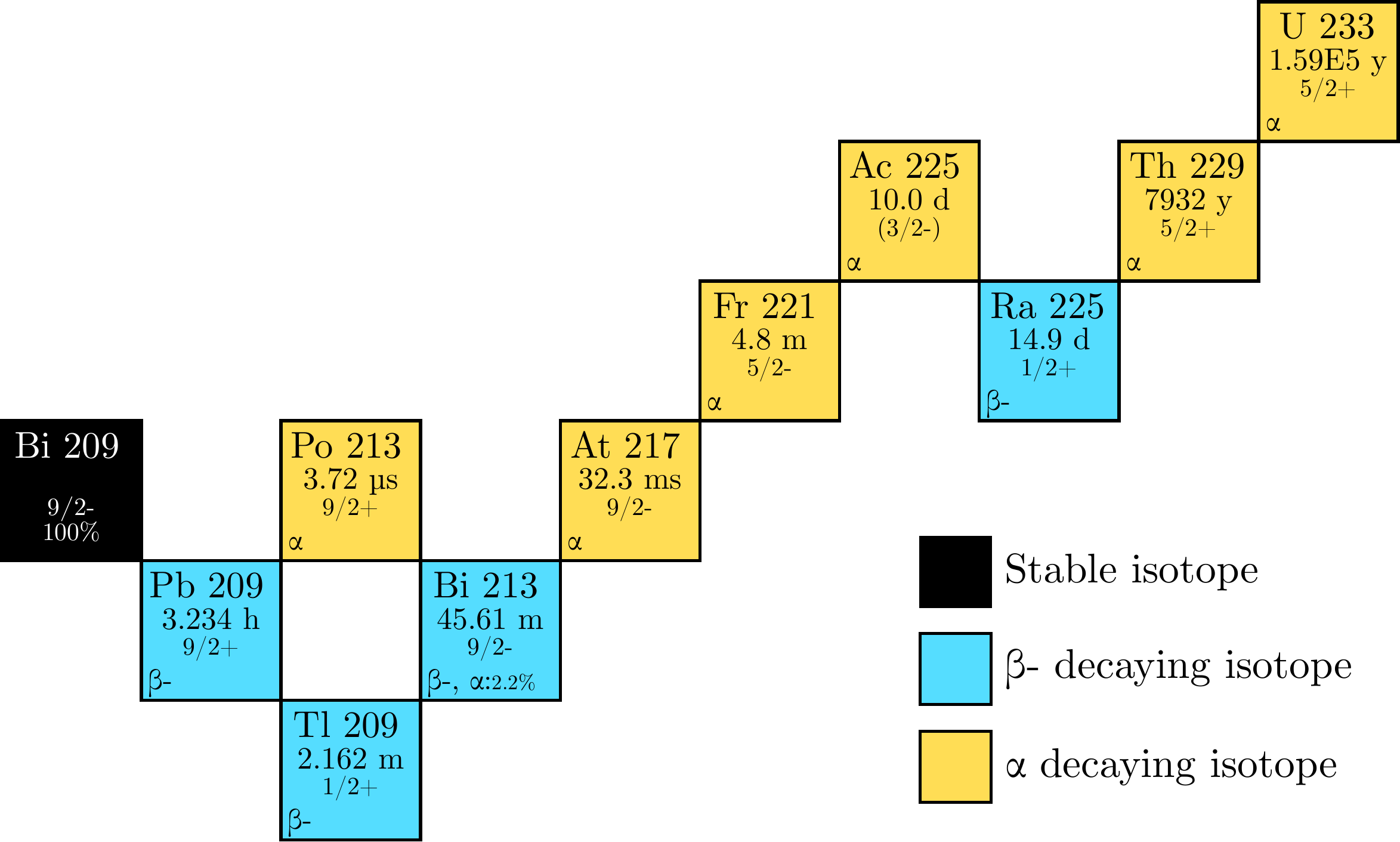}
  \caption{The~$^{233}$U decay chain. The decay mode with possible branching, half-life and ground-state spins are given for each isotope. Decay data is from reference~\cite{nndc-website}.}\label{fig:U233_decay_chain}
\end{figure}

The decay chain of~$^{233}$U is shown in Fig.~\ref{fig:U233_decay_chain}. Due to the long half-life of~$^{229}$Th (7.9$\cdot 10^3$ years~\cite{Varga2014}), it starts to build up in the source material and the source becomes activated with all isotopes in the decay chain. In particular with an aged source,  in addition to~$^{229}$Th, alpha recoils of all isotopes can be expected to be produced that are daughters of alpha-active mothers.

Despite the small branching ratio to the isomeric state, using~$^{233}$U as a production mechanism for a~$^{229}$Th source should be a viable method to produce a high flux if the efficiency can be optimized. This approach has already been studied previously at the IGISOL facility~\cite{Sonnenschein2012}. Several iterations of the gas cell were implemented and tested in order to improve the extraction efficiency of the~$^{229}$Th recoils. The best efficiency reported, however, was only 1.6\%, attained with ion extraction using an electron emitter that created a strong electric field inside the gas cell effectively ``pulling'' the positively-charged ions out.

The focus in the current work has shifted to the sources themselves and developments have continued with the study of a new~$^{233}$U source in addition to the existing sources. A more careful study of the recoil efficiency, defined as the number of recoils leaving a source divided by the total number of recoils, has been performed as this critical factor is sensitive to many parameters. In this article the results of spectroscopic analyses of~$^{233}$U sources for their characterization and determination of recoil efficiency are presented. The characterization has been done using the alpha- and gamma radiation emitted directly from the sources and via the radiation emanating from alpha recoils collected on implantation foils. Also, the source structure and composition have been studied in a series of Rutherford backscattering spectrometry (RBS)
measurements.

\section{Description of the sources}

\subsection{Thickness limit for a source layer}

As the recoil energy of the daughter nucleus~$T_{\mathrm{D}} = m_{\mathrm{\alpha}}/(m_{\mathrm{\alpha}}+m_{\mathrm{D}}) Q_{\mathrm{\alpha}}$ is low, only~$\sim84$~keV, the source characteristics need to be very well controlled in order to obtain the maximum number of recoils. If the source is too thick, has poor homogeneity, contains a large number of impurities, has contaminant surface layers, or the substrate roughness is not controlled, significant recoil losses can already occur in the source material. Therefore it is important to characterize and measure the efficiency of the sources.

The limit for the source thickness can be taken as the distance in which the recoils from the lowest~$^{233}$U atomic layer are stopped inside the source material. In order to have a quantitative number for this limit, the software package Stopping and Range of Ions in Matter (SRIM)~\cite{Ziegler2010} was used to calculate the stopping ranges of~$^{229}$Th ions in pure~$^{233}$U metal,~$\ce{UO2}$ and in~$\ce{UF4}$. The calculations predict stopping ranges of 19~$\mathrm{\upmu g/cm²}$~($\sim$10~nm), 16~$\mathrm{\upmu g/cm²}$~($\sim$15~nm), and 14~$\mathrm{\upmu g/cm²}$~($\sim$21~nm), respectively, with straggling of $\pm12~\mathrm{\upmu g/cm²}$~($\pm6$~nm), $\pm9~\mathrm{\upmu g/cm²}$~($\pm7$~nm) and $\pm7~\mathrm{\upmu g/cm²}$~($\pm9$~nm) as defined by the square root of the position variance~$\sqrt{\langle(\Delta x)^2\rangle}$.

\subsection{Energy distribution of recoils}

\begin{figure}
  \centering
  \includegraphics[width=\columnwidth]{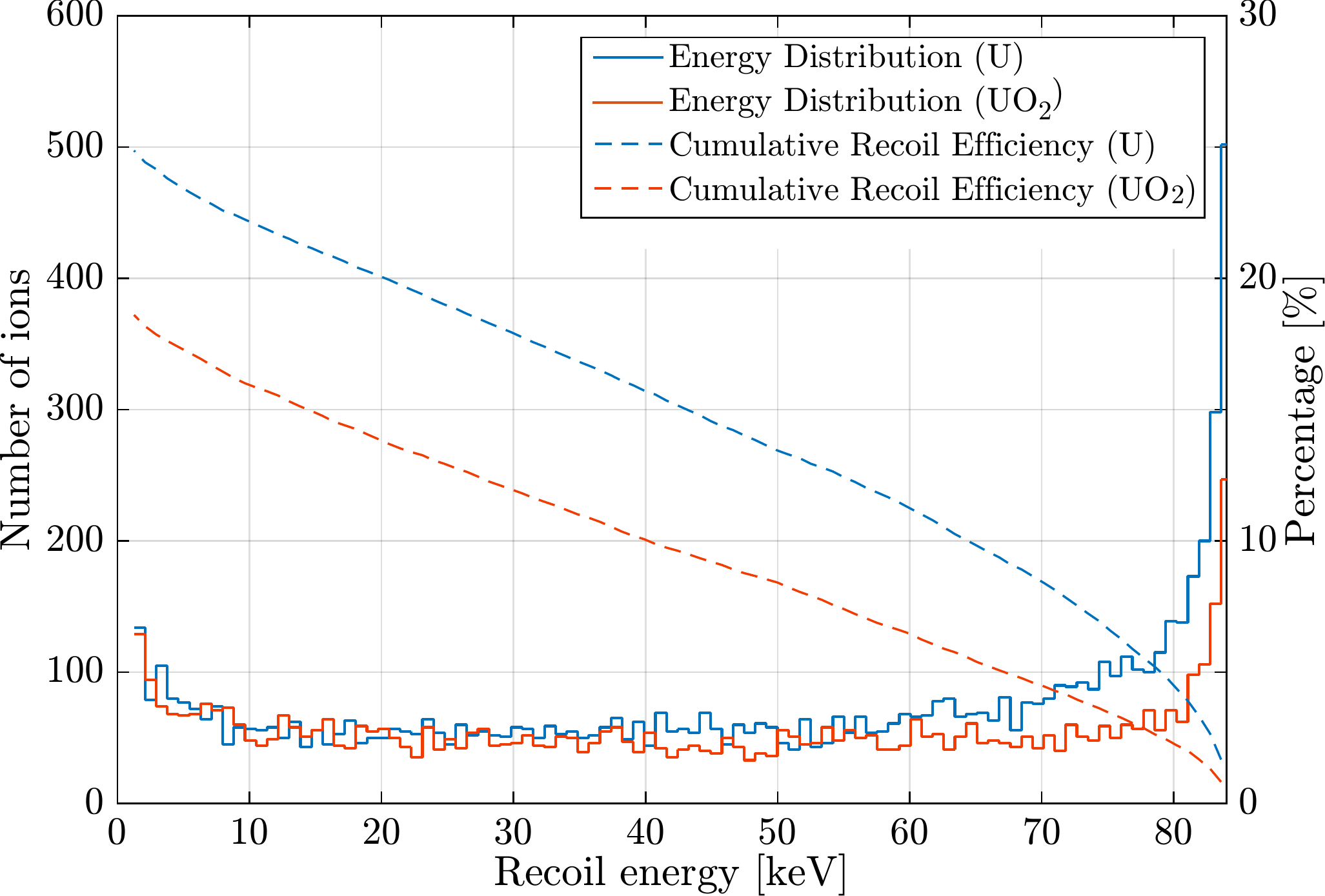}
  \caption{The simulated~$^{229}$Th recoil energy distribution (histogram lines) and recoil efficiency as a reverse cumulative percentage (dashed line) when the~$^{229}$Th recoils are coming out of a pure~$^{233}$U source or an oxidized~UO$_2$ source.}\label{fig:th_recoil_e_dist}
\end{figure}

The TRIM Monte-Carlo processor of the SRIM software package was used to estimate the recoil energy and the recoil efficiency from a pure~$^{233}$U layer and oxidized~$^{233}$U assuming UO$_2$. Both simulations were done with a~$^{233}$U areal density of 75$\cdot10^{15}$ at./cm$^2$, approximately matching the~$^{233}$U layer in the thinner source analyzed in this work. The energy distribution of the~$^{229}$Th recoils, which exit the source material, is shown with the histogram plots in Fig.~\ref{fig:th_recoil_e_dist}, illustrating a somewhat constant distribution between energies of 0 to 60 keV after which the distribution increases exponentially. This increase is the result of the very top layer of atoms of the source decaying in which there is very little material for the recoils to lose energy.

The recoil efficiency is shown also in Fig.~\ref{fig:th_recoil_e_dist} with dashed lines plotting the reverse cumulative percentage of the total number of decays. If the recoils with an energy above 10 keV are considered, a recoil efficiency of approximately 22\% is expected for a pure~$^{233}$U layer and 16\% for~UO$_2$. Because the oxidation of uranium is difficult to avoid in practice, 16\% can be taken as the best possible recoil efficiency attainable with a source having a~$^{233}$U density of 75$\cdot10^{15}$ at./cm$^2$.

\begin{figure*}
  \includegraphics[width=\textwidth]{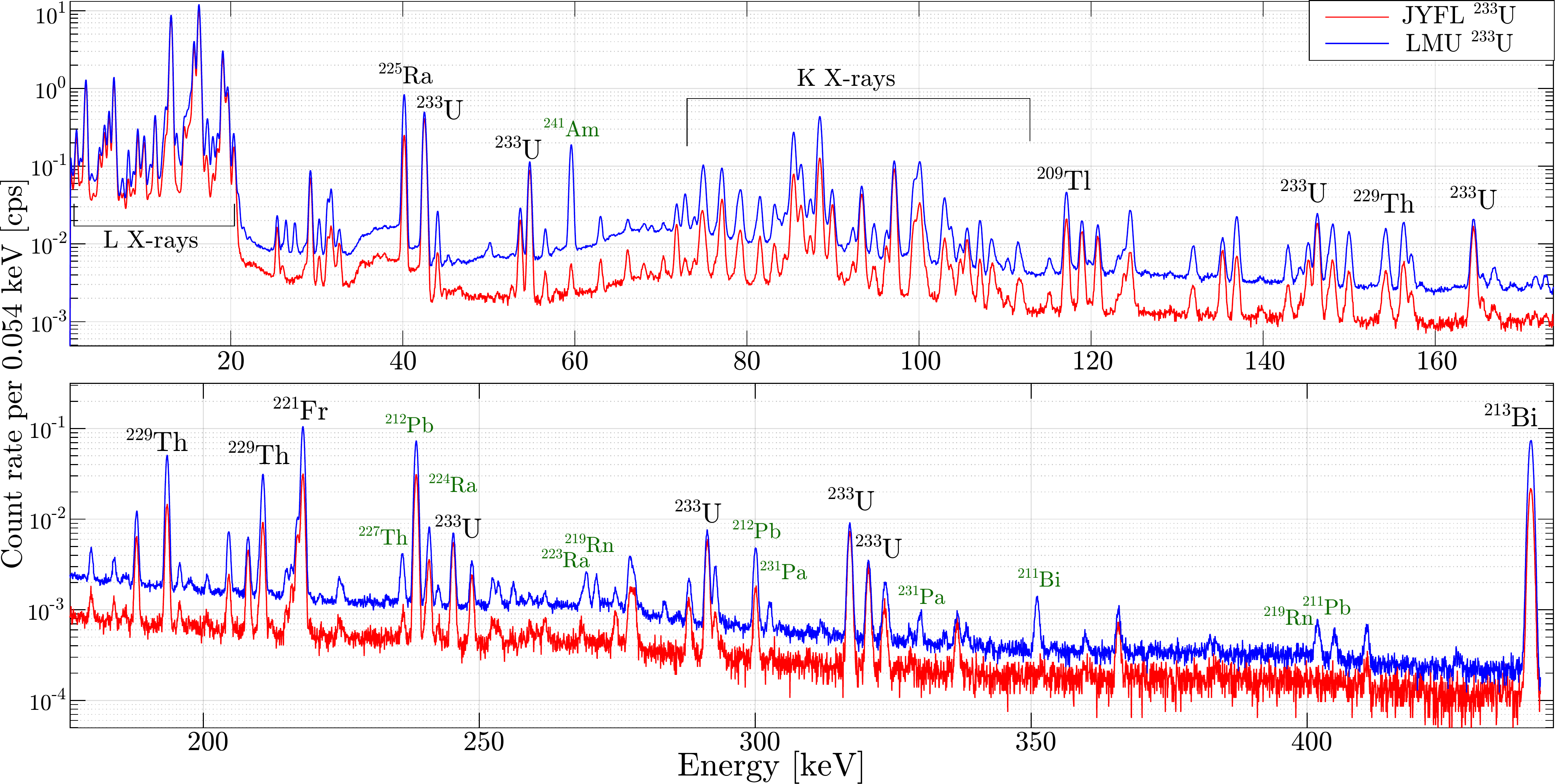}
  \caption{Direct gamma spectra of the LMU and JYFL~$^{233}$U sources between the energy range 5 keV to 500 keV. All the identified~$^{233}$U decay chain isotopes are marked with black labels and other identified isotopes with green labels.}\label{fig:U233_direct_gamma}
\end{figure*}

\subsection{Specifications of the studied sources}

The spectroscopic analysis was performed on two different types of $^{233}$U source. One was provided by the Ludwig Maximilian University of Munich~\cite{Wense2015} and hereafter will be referred to as the LMU source. The source was made by evaporating uranium tetrafluoride on an aluminium backing until an activity of 230~kBq was accumulated within a circular area of 20 mm in diameter. Using the known density of $^{233}$UF$_{4}$, 6.7 g/cm$^{3}$~\cite{crc-book}, a source thickness of $\sim$420~nm was inferred.

Another~$^{233}$U source, originally manufactured by McGill University through neutron capture of~$^{232}$Th, is a set of stainless steel foils on which~$^{233}$U was electrodeposited in a thin layer~\cite{Tordoff-thesis}. Henceforth, these sources will be referred to as the JYFL sources. A total activity of 2.3~MBq of~$^{233}$U is available, distributed over six strips with dimensions 25~mm~$\times$~70~mm, and sixteen strips with dimensions 12~mm~$\times$~70~mm; in total about 240~cm$^2$. Assuming the uranium is evenly deposited and oxidized to~$\ce{UO2}$ (10.97~g/cm$^{3}$~\cite{crc-book}) due to unavoidable exposure to air, a thickness of 30~nm can be deduced, albeit the exact thickness will depend on the crystal structure and oxidation state of the uranium which are not known. Table~\ref{tab:source_properties} provides a summary of the source specifications along with photographs.

\begin{table}
  \begin{tabular}{r | l | l }
    \toprule
    & \includegraphics[width=0.18\textwidth]{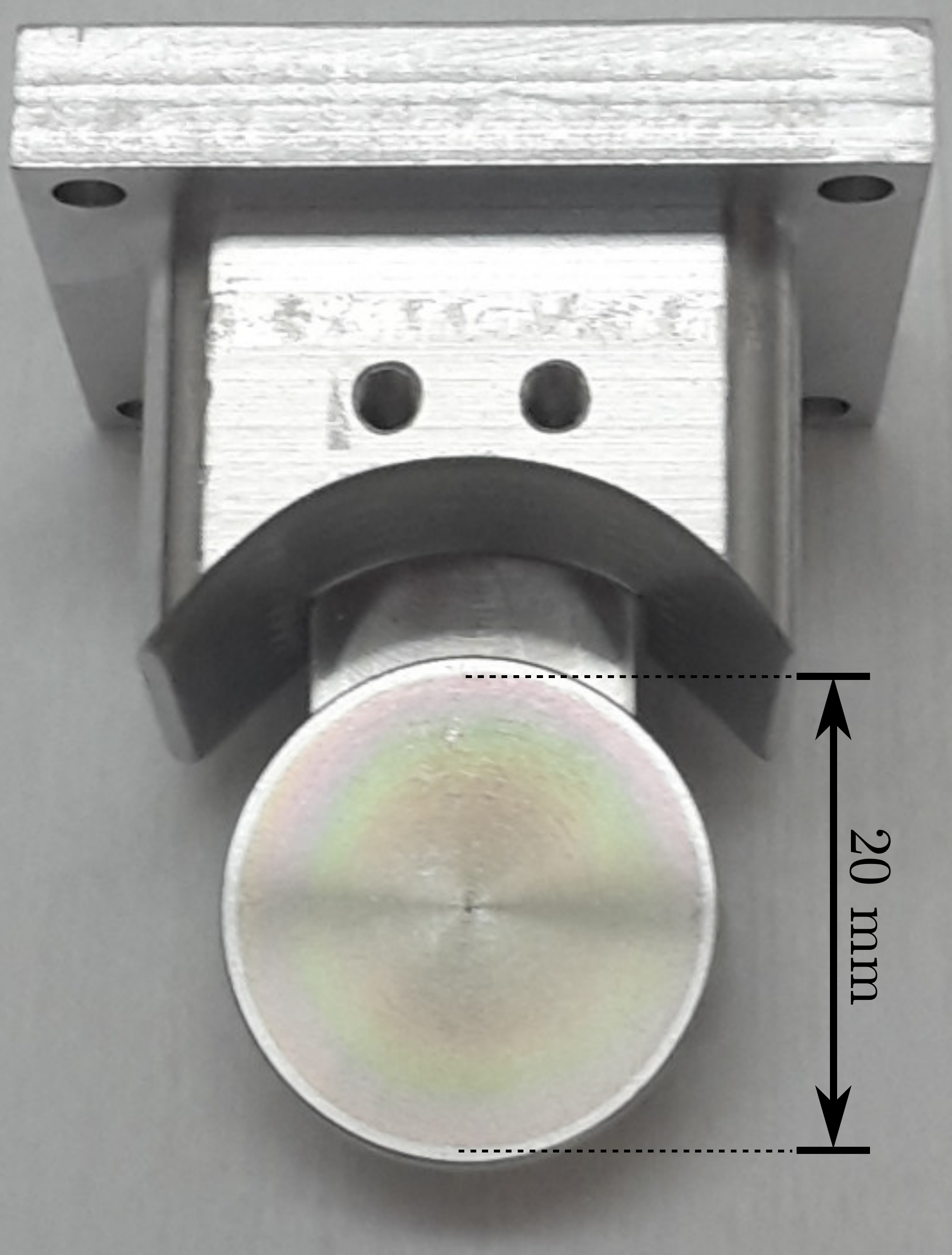} & \includegraphics[width=0.14\textwidth]{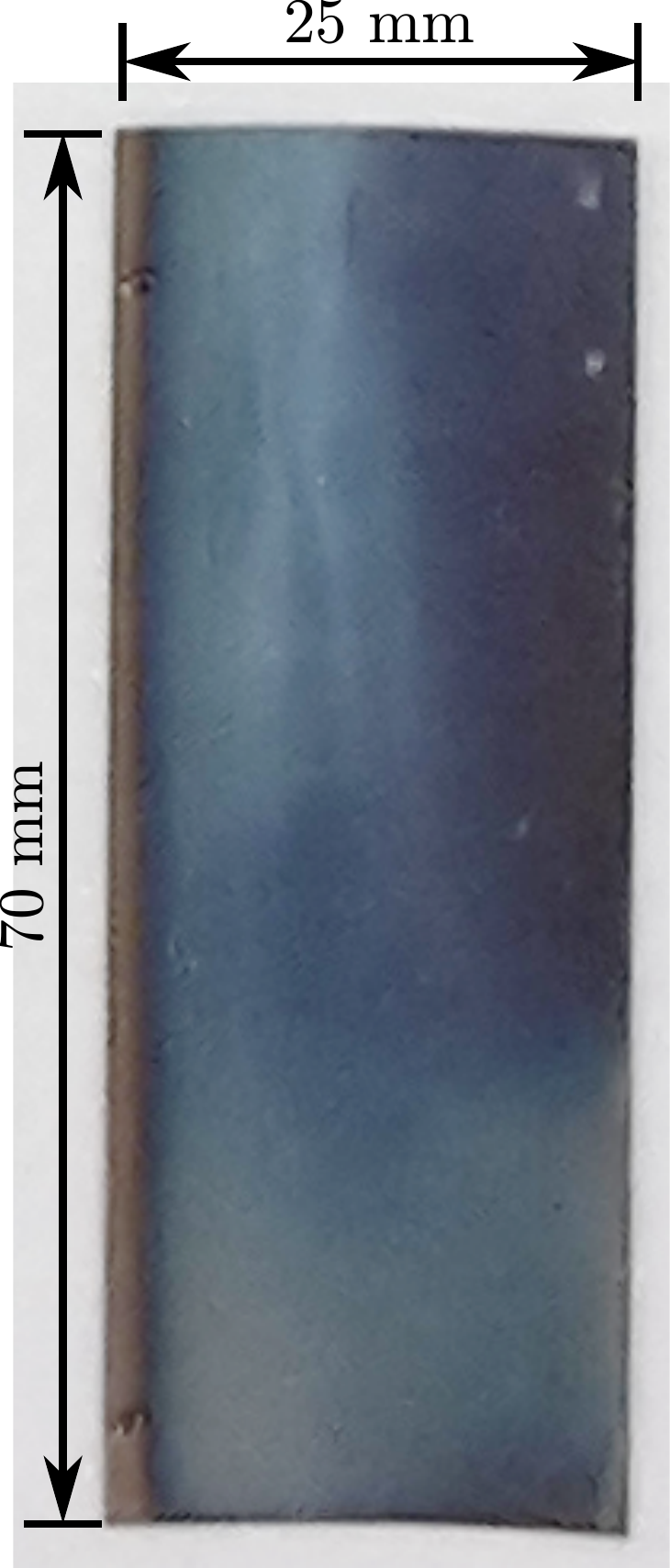}\\ \hline
    \textit{Source}& LMU& JYFL\\
    \textit{Composition} & $\ce{UF_4}$ & $\ce{UO2}$\\
    \textit{Backing} & Aluminium & Stainless steel \\
    \textit{Deposition} & Evaporated & Electroplated \\
    \textit{Activity} & 73.2~kBq/cm$^2$ & 10.2 kBq/cm$^2$\\
    \textit{Areal density} & $530\cdot10^{15}$ at./cm$^2$ & $74\cdot10^{15}$ at./cm$^2$ \\
    \textit{Geometry} & disc dia. 20 mm  & 25 mm $\times$ 70 mm\\    
    \textit{Mass density} & 0.27 mg/cm$^2$ & 0.03 mg/cm$^2$\\
    \textit{Thickness} & $\sim$420 nm & $\sim$30 nm \\
    \textit{Total activity}& 230 kBq & 2.3 MBq \\
    \textit{Total area} & 3.1 cm$^2$& 240~cm$^2$ \\
    \bottomrule
  \end{tabular}
  \caption[Physical properties of the studied $^{233}$U sources.]{Physical properties of the studied $^{233}$U sources. Thicknesses are given assuming~$\ce{UF4}$ and $\ce{UO2}$ for the LMU and JYFL sources, respectively. Total activity and area are given for all the available JYFL source strips.}\label{tab:source_properties}
\end{table}

\section{Source characterization by gamma-ray and alpha radiation measurement}
\subsection{Direct source radiation}

In order to verify the strength of the two sources and to measure the amount of activity of the daughter isotopes, a gamma-ray spectroscopic measurement was performed in a low-background counting station. Figure~\ref{fig:U233_direct_gamma} highlights the X-ray and gamma-ray peaks identified in the lower energy region of the spectra, from a few keV to above 400 keV for the two sources. The overall background of the JYFL source is lower reflecting a lower source strength from a single strip as well as reduced daughter and contaminant activities.

Following peak identification, activities of $^{233}$U and $^{229}$Th were determined by integration of selected gamma-ray peaks while taking into account the branching ratios and detection efficiency. In this manner, the activity for the LMU source was measured to be 230(10) kBq of $^{233}$U and 1.2(0.2) kBq of $^{229}$Th, while a single (large area) JYFL strip resulted in activities of 200(10) kBq and 350(40) Bq, respectively. The age of the~$^{233}$U source material of the LMU and JYFL sources was estimated to be 60(11) years, and 20(3) years, respectively, through the equation $t = A(^{229}\mathrm{Th})/\left(\lambda(^{229}\mathrm{Th})A(^{233}\mathrm{U})\right)$. These age estimations are purely from the measured activities and do not take into account the~$^{229}$Th that has not stayed within the uranium. Still, the ages match roughly with those expected for the source materials, and the 60(11) year estimate matches also the previous estimation of 45(5) years~\cite{Wense2015}.

By using a silicon charged-particle detector (Ortec U-017-300-500) with an intrinsic resolution of 17 keV, 500~$\upmu$m depletion depth and 300~mm$^2$ detection area, a direct measurement of the alpha decay from the two sources was compared. By measuring the alpha-particles coming out of the source, the activities of $^{233}$U and its decay chain isotopes as well as radiogenic impurities can be obtained in a similar manner to the gamma-ray measurement. Alpha particles also reveal possible surface characteristics from the straggling as they pass through the material.

\begin{figure}
  \centering
  \includegraphics[width=\columnwidth]{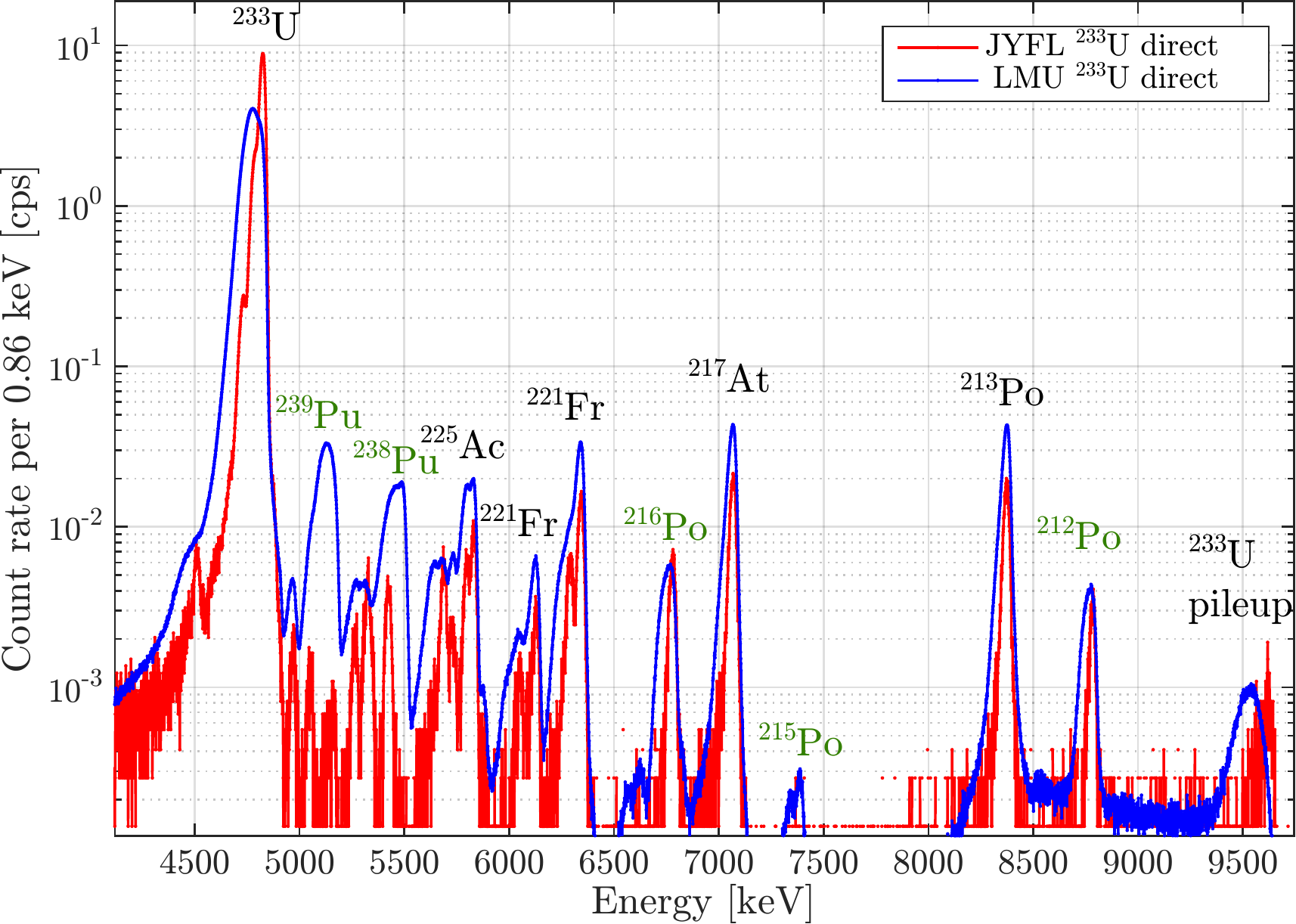}
  \caption{Direct alpha spectra of the~$^{233}$U sources. All the identified~$^{233}$U decay chain isotopes are marked with black labels and other identified isotopes with green labels. The larger alpha peak widths of the LMU spectrum are due to a thicker source layer.}\label{fig:U233_direct_alpha}
\end{figure}

Figure~\ref{fig:U233_direct_alpha} shows the resulting alpha spectra when the detector was approximately 12~cm away from the sources in a rough vacuum environment, again highlighting a direct comparison of the two sources. In addition to the decay chains of $^{233}$U and $^{232}$U, lines of $^{238}$Pu and $^{239}$Pu are visible, the latter decays particularly noticeable in the LMU source. A direct measure of the alpha decays from $^{229}$Th is not possible due to overwhelming background from the alpha line associated with the decay of $^{233}$U. Finally, the typical alpha-decay peak width of the LMU source was measured to be~100 keV, while alpha-width of the JYFL source was measured to be~30 keV, close to the nominal detector resolution of about 20 keV. The broadening of the peak structure reflects a combination of the source thickness as well as possible contaminants on top of the source. The thickness of source material was also observed through the increased broadening of alpha peaks when the sources were tilted at an angle with respect to the silicon detector.

\subsection{Source characterization by foil implantation}

In order to measure the rate of alpha recoils, including~$^{229}$Th, released from the LMU and JYFL~$^{233}$U sources, a foil implantation experiment was performed. To collect the recoils, aluminium foils were mounted at a distance of approximately 2 mm from the $^{233}$U source surface in a chamber connected to a scroll vacuum pump. A vacuum of $\sim$10$^{-1}$ mbar was required during the implantation process as the recoil stopping in air would otherwise be significant. The implantation was performed over a duration of approximately 12.6 days for the JYFL source and 46.3 days for the LMU source.

\begin{figure*}
  \centering
  \includegraphics[width=0.9\textwidth]{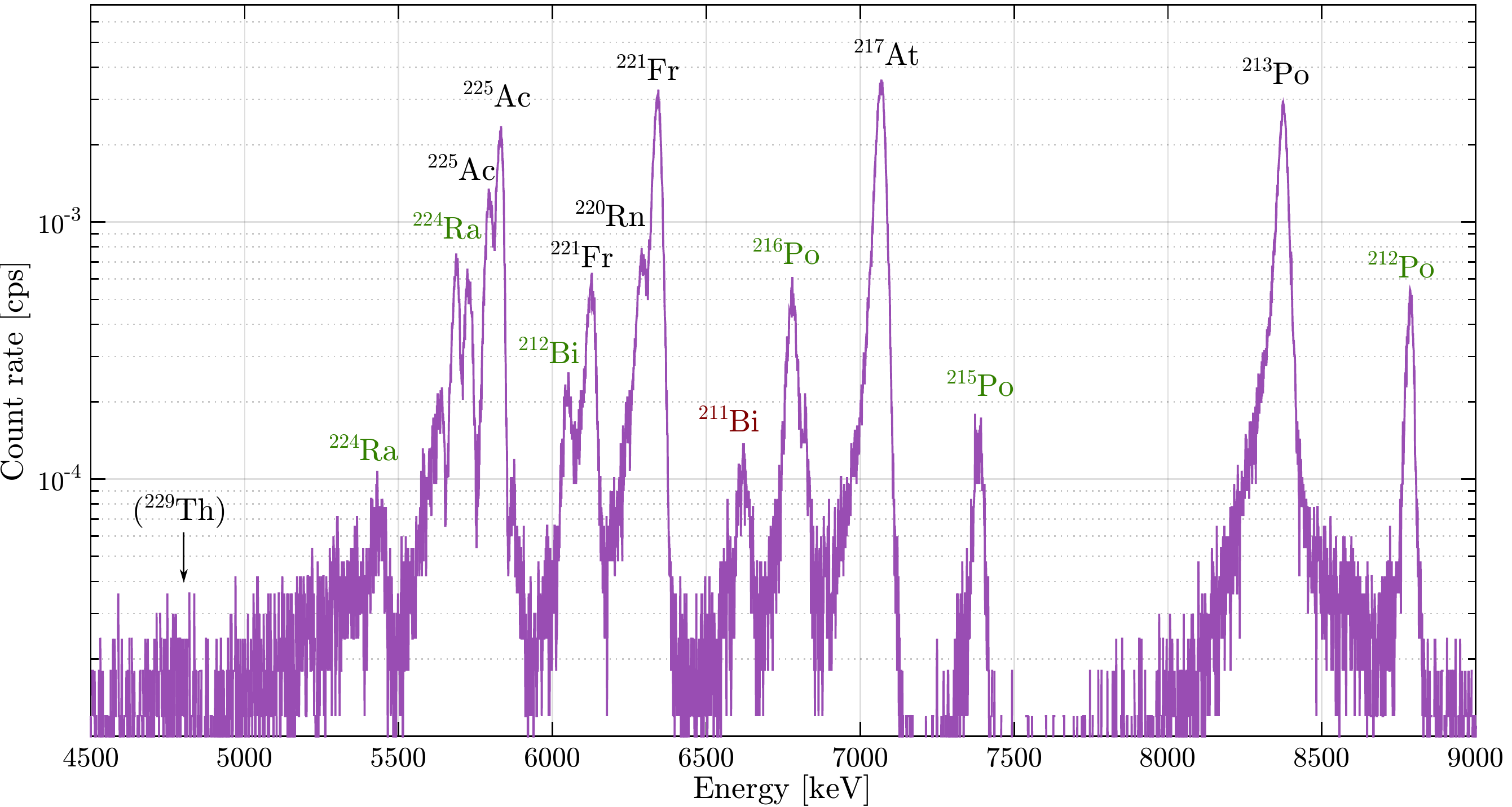}
  \caption{The alpha spectrum of the Al foil following a 4-day implantation using the LMU~$^{233}$U source. Several isotopes belonging to the~$^{233}$U decay chain (black labels) and others from~$^{232}$U and possibly~$^{227}$Ac chains are identified. Unfortunately, due to the long half-life of~$^{229}$Th, its alpha decays were not detected above background.}\label{fig:munic_foil_alpha_labeled}
\end{figure*}

Figure~\ref{fig:munic_foil_alpha_labeled} shows the alpha spectrum of the foil implanted using the LMU source, measured with the Si detector for 4 days. A clear activity of isotopes from the $^{233}$U decay chain can be seen. As the alpha decays from the $^{229}$Th recoil ions are not visible due to the long half-life of $^{229}$Th and background counts, we can only assign an upper limit of about 30000 recoil ions per second implanted or about 13\% of the total activity  of the LMU source. Because the~$^{229}$Th activity was not directly visible in the alpha spectrum, the implanted foils were measured in the low-background gamma radiation detection station in the hope of detecting the 193.5 keV and 210.9 keV gamma lines as they have a reasonable branching of 4.3\% and 2.8\%, respectively~\cite{Jain2009}.

\begin{figure*}
  \centering
  \includegraphics[width=0.9\textwidth]{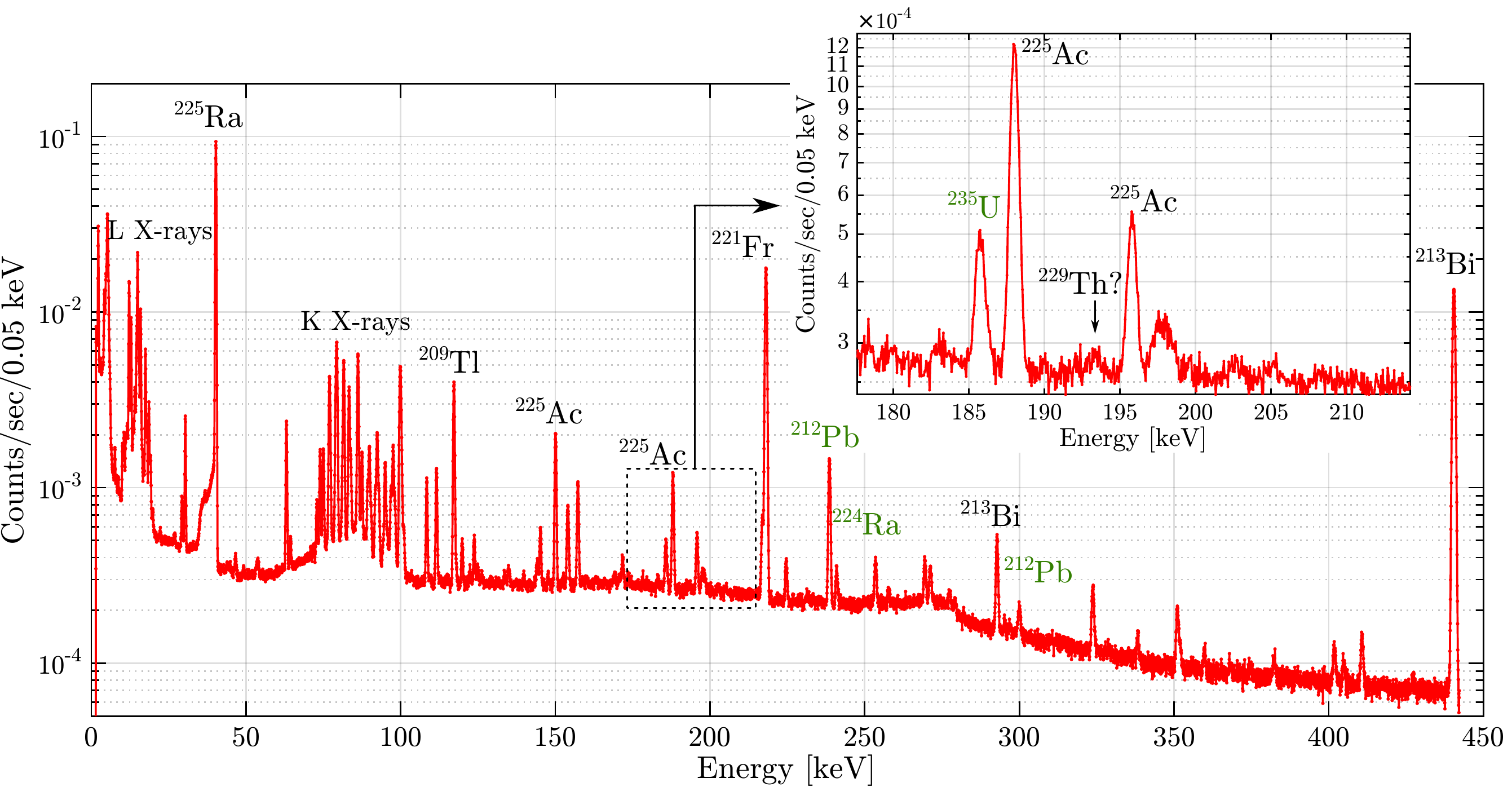}
  \caption{Gamma spectrum of the Al foil irradiated for a period of 46.3 days by the LMU~$^{233}$U source. All identified isotopes belonging to the~$^{233}$U decay chain are labeled in black and other assignments are labeled in green. The enlarged region shows the expected position of the ~$^{229}$Th peaks of which the 193.5 keV is barely visible above background.}\label{fig:LMM_foil_gamma}
\end{figure*}

Unfortunately, even with the long implantation time using the LMU source and a 30-day measurement in the low-background gamma station, the 193.5~keV gamma line was barely visible above background, as seen in the gamma-ray spectrum shown in Fig.~\ref{fig:LMM_foil_gamma}. Nevertheless, the integration of the area of the 193.5~keV peak was possible suggesting a recoil ion efficiency of~$\sim$3\%, a factor of two lower than the estimate provided by L.v.d. Wense et al.~\cite{Wense2015} for the same source. This direct measurement of the~$^{229}$Th recoil efficiency can be taken only as an upper limit because the gamma line is not clearly apparent.

Similar measurements of recoil ion implantation and gamma-ray analysis was performed using the JYFL source. Worryingly, the gamma line was not at all visible above the background, which can at least partly be associated with a shorter implantation time.

As the~$^{229}$Th activity was not directly discernible in the implantation foils, a determination of the recoil efficiency of the sources had to rely on the implantation of~$^{225}$Ra daughter isotopes. Similar to~$^{229}$Th recoils of~$^{233}$U, the~$^{225}$Ra are recoiling out of the sources due to the decay of~$^{229}$Th which has accumulated into the sources after manufacturing. The detection of~$^{225}$Ra is more readily done due to its significantly lower half-life than~$^{229}$Th. The amount of~$^{225}$Ra was determined from its decay gamma-ray peak at 40 keV as well as from the amount of~$^{225}$Ac, the decay product of~$^{225}$Ra. Using the determined~$^{229}$Th activity from the direct gamma radiation measurements, a recoil efficiency of $^{225}$Ra of 3\% for the JYFL source and 2\% for the LMU source can be calculated. Assuming that the ratio between the recoil implantation rates of~$^{229}$Th and~$^{225}$Ra is the same as the detected activity ratio of~$^{233}$U and~$^{229}$Th in the sources, the deduced recoil efficiency for~$^{225}$Ra can be taken as an approximate number for the~$^{229}$Th recoil efficiency. The details of the recoil efficiency determination can be found in reference~\cite{Ilkka-thesis}. It should be noted that the efficiency may be underestimated if the decaying~$^{229}$Th distribution reaches deeper than the~$^{233}$U distribution or overestimated if~$^{229}$Th atoms are implanted into a contaminant layer on top of the~$^{233}$U layer.

Still, considering the low daughter recoil efficiency and the low detected amount of~$^{229}$Th, which should have been easily directly detectable if efficiencies would have been optimum, it is obvious that the recoil efficiency needs to be improved, especially for the JYFL source where the simulated recoil efficiency is estimated to be about 16\% as was shown by the TRIM simulations.

The measured recoil efficiency of 2\% for the LMU source is about a factor of two to three lower than that estimated for the same source (and also using the daughter activity) by~L.v.d. Wense et al.~\cite{Wense2015}. The value presented here is close to the recoil efficiency of~$\sim$1.6\% which was estimated by L.v.d. Wense~\cite{Wense-thesis} for 360~nm of~$^{233}$UF$_4$ material using a SRIM calculation and a model for isotropic emission of alpha recoils. The higher recoil efficiency estimate was attributed to possible channeling of the recoils in a crystallized surface layer. With thicker sources it is clear that the recoil efficiency cannot be improved significantly and therefore the approach of L.v.d. Wense et al.~\cite{Wense-thesis, Wense2016} to use a thinner 290~kBq source that was deposited on a Ti-sputtered Si wafer of 100 mm in diameter, is the correct approach, resulting in a significantly improved recoil efficiency of~$\sim$35\%.

\section{Source characterization by Rutherford backscattering spectrometry}\label{subsec:RBS}

\subsection{Measurement of RBS spectra}
\begin{figure}
  \centering
  \includegraphics[width=\columnwidth]{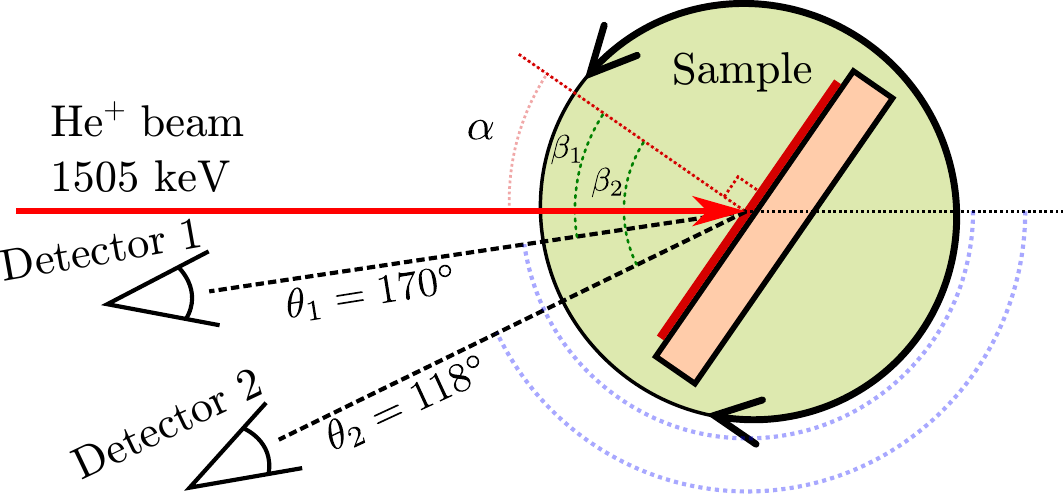}
  \caption{The measurement geometry for the RBS measurement of the~$^{233}$U sources using IBM geometry with~$\beta - \alpha + \theta = 180\degree$.}\label{fig:rbs_geometry}
\end{figure}

Due to the low recoil efficiencies of the~$^{233}$U sources as shown by the implantation measurements, the RBS analysis technique~\cite{Putkonen2005} was used to investigate the composition of the source layers. A 1505 keV He beam was produced with the 1.7 MV NEC Pelletron accelerator located within the JYFL Accelerator Laboratory. The He beam was directed to the sources which were mounted inside a dedicated RBS measurement chamber. Both of the sources were characterized using IBM geometry in which the beam incident angle~$\alpha$ and exit angle of detected particles, the angle~$\beta$, are kept on the same plane as the sample surface normal. While performing sample tilting so that~$\beta - \alpha + \theta = 180\degree$, as depicted in Fig.~\ref{fig:rbs_geometry}, the backscattered He beam was measured with two detectors at backscattering angles of~$\theta_1 = 170$\degree~and~$\theta_2 = 118$\degree.

\begin{figure*}
  \centering
  \includegraphics[width=0.9\textwidth]{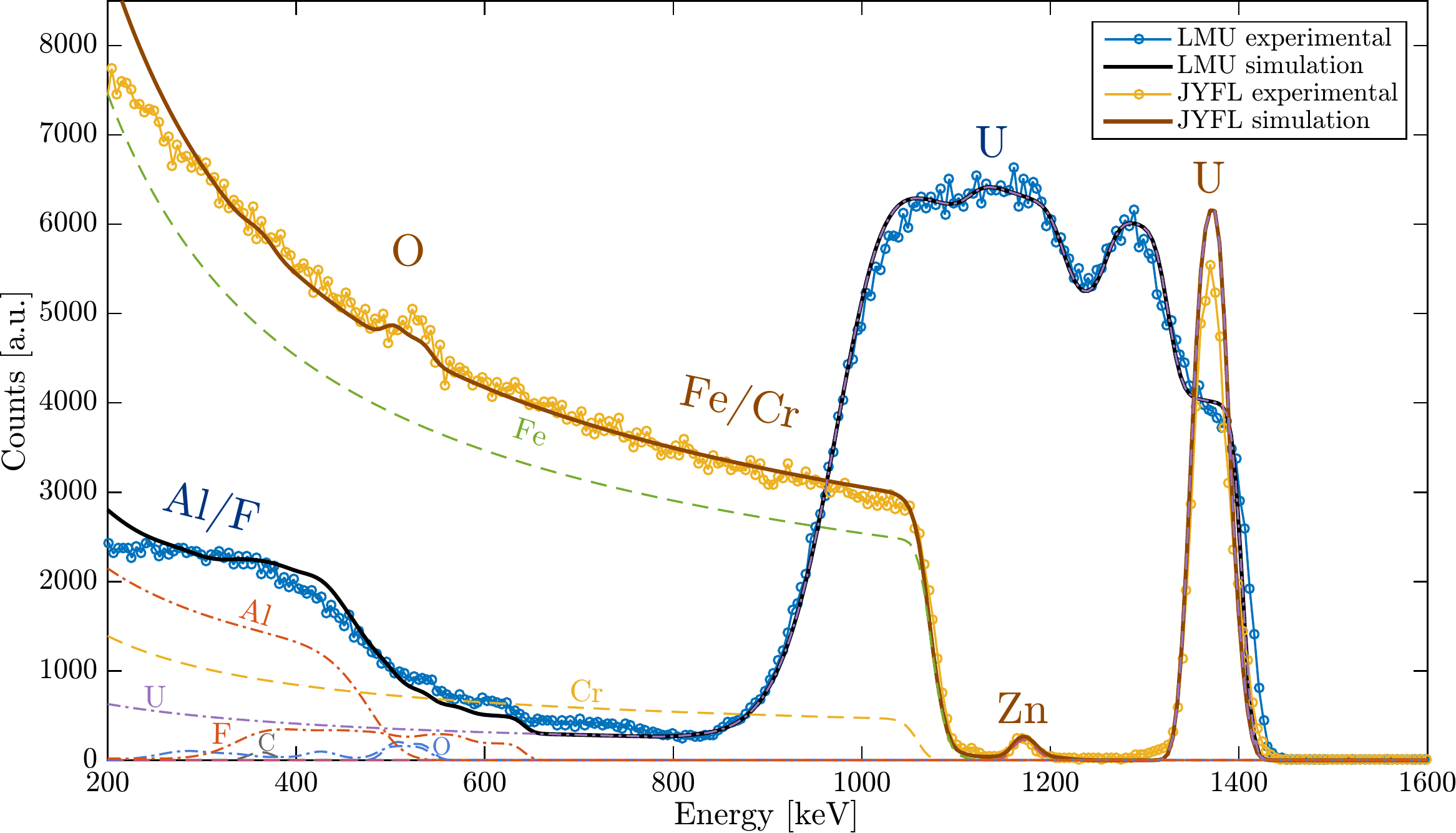}
  \caption{The RBS and Simnra simulation spectra of the LMU and JYFL~$^{233}$U sources. The individual elements in the simulation are plotted with dashed lines for the JYFL source simulation and dash-dotted lines for the LMU source simulation.}\label{fig:rbs_compare}
\end{figure*}

The experimental RBS spectra of the LMU and JYFL sources measured with detector 1 at a tilt angle~$\alpha = 0$\degree~are shown in Fig.~\ref{fig:rbs_compare}. The~$^{233}$U is visible in both spectra as the backscattering with the highest energy. In the JYFL source, in which the~$^{233}$U layer is much thinner ($\sim${}$74\cdot10^{15}$ at./cm$^{2}$) than in the LMU source, the uranium appears as a sharp peak at $\sim$1400 keV with the stainless steel substrate as the continuous structure starting from about 1100 keV. The interesting feature of the spectrum is the peak between the uranium and the substrate, which has been identified as a thin layer of zinc. Also, there is clear peak in the signal at around 500 keV indicating an oxide layer.

Because the~$^{233}$U layer is much thicker (${\sim}530\cdot10^{15}$ at./cm$^{2}$) in the LMU source, the corresponding uranium peak is much wider showing also structural variations in the source layer. As the substrate is aluminium its edge is found at lower energy than for stainless steel. As a light element, the fluorine in the source layer is also found in the low-energy region. 

\subsection{Simulation of the RBS spectrum}

For both spectra the elemental composition was determined by using the RBS simulation software Simnra~\cite{simnrareport}. Figure~\ref{fig:rbs_compare} shows the simulation results as solid lines with individual elements as dashed lines for the JYFL source and dash-dotted lines for the LMU source. In order to get a good fit to the experimental data, the source structures were defined as several separate layers and the layer composition was then optimized to obtain the best fit. As the sources were measured with two detectors at several tilt angles ($\alpha = 0\degree, 5\degree, 10\degree$ and~$15\degree$), the layer structure was iteratively optimized until the simulation consistently matched to all measured data giving more confidence to the simulation.

\begin{figure}
    \centering
    \subfloat[\label{fig:simnra_rults_LMU}]{\includegraphics[width=0.9\columnwidth]{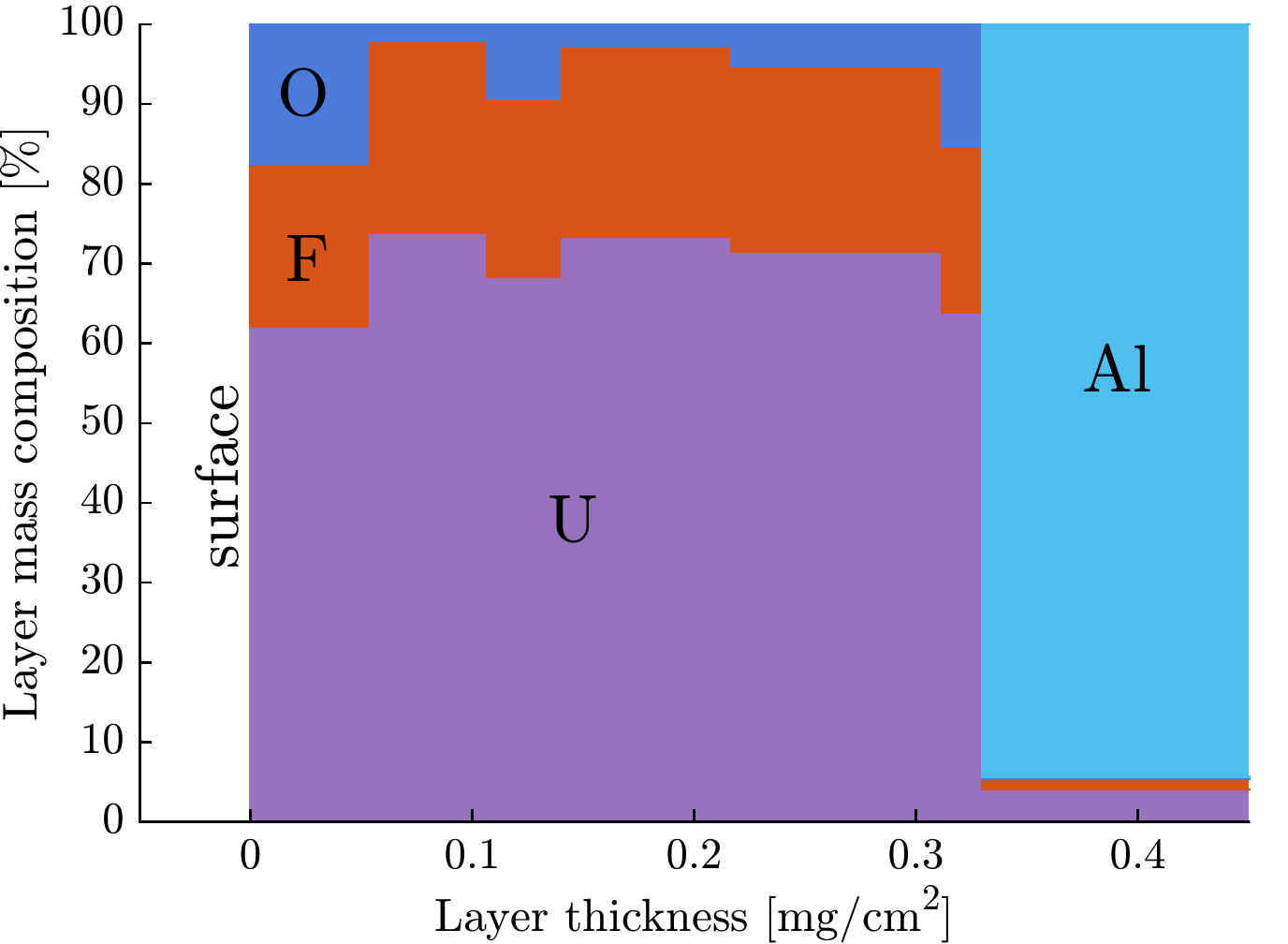}}
    \vspace*{\fill}
    \centering
    \subfloat[\label{fig:simnra_rults_JYFL}]{\includegraphics[width=0.9\columnwidth]{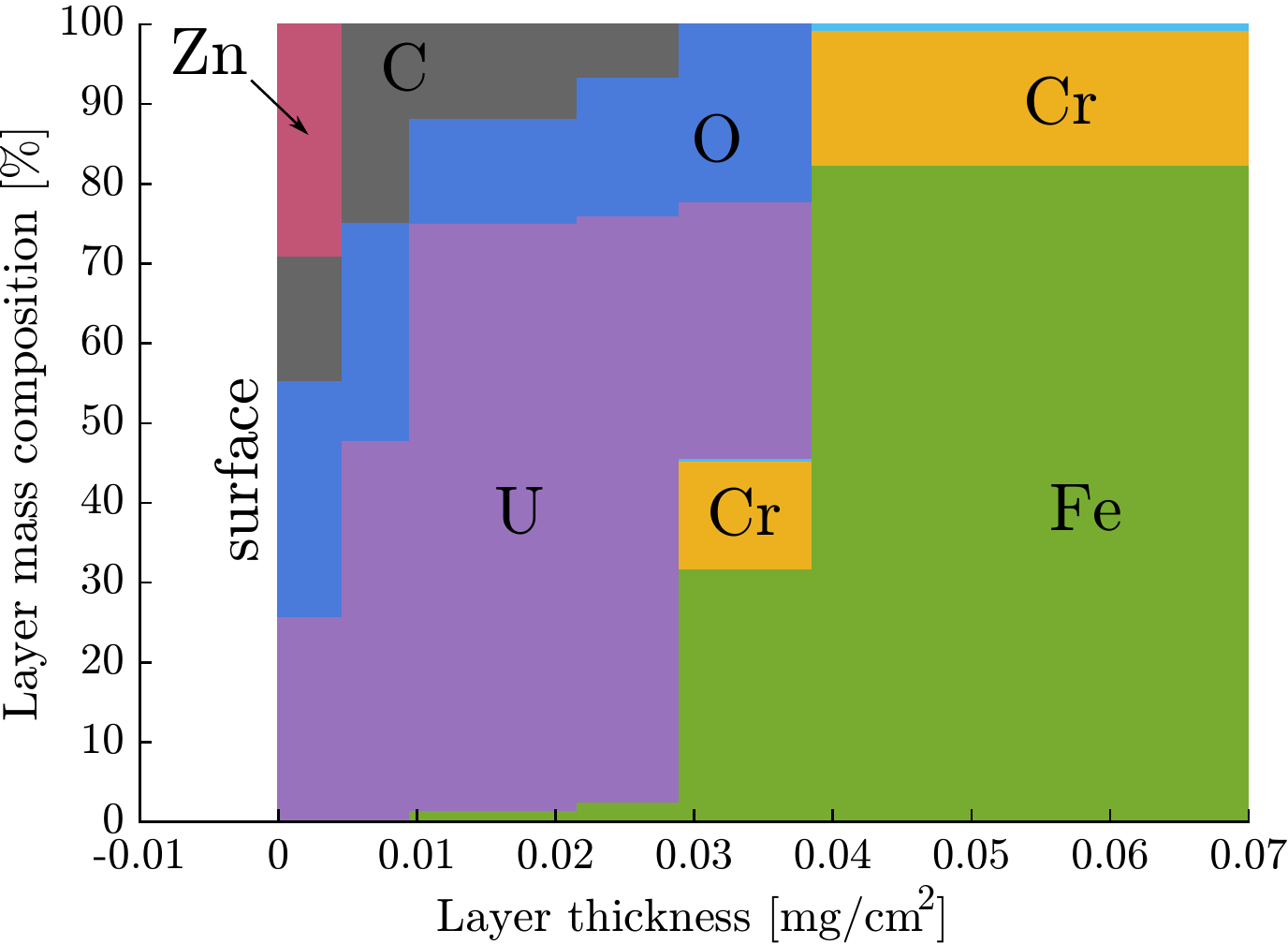}}
    \par\vfill
  \caption{The elemental mass composition as a function of layer depth of the~$^{233}$U LMU (a) and JYFL (b) sources according to the Simnra simulation.}\label{fig:simnra_rults}
\end{figure}

The elemental composition of sources as a mass percentage according to the simulations is shown in Fig.~\ref{fig:simnra_rults} as a function of depth. The surface layer is the leftmost layer. In the LMU source the layer composition has been assumed to be~$\ce{UF4}$ into which oxygen was added in order to match the uranium peak width. As indicated by the experimental spectrum, the extra material in the~$\ce{UF4}$ is assumed to be a light element, and therefore modeled as oxygen. However, we were not able to exactly identify the material embedded with the~$\ce{UF4}$ and the peak width might, for example, result from a higher degree of fluorination of the uranium or indication of impurities being introduced during the evaporation process. These impurities might also explain the density variations in the uranium peak. 

The low-energy tail of the uranium peak, which is expected to be caused by distinct pieces of~$\ce{UF4}$ being on top of the~$\ce{UF4}$ layer, is simulated by adding a small amount of~$\ce{UF4}$ to the aluminium substrate. The amount of~$^{233}$U itself is 640$\cdot10^{15}$~at./cm$^{2}$ according to the simulations, matching reasonably well with the estimated thickness as derived from the activity. Finally, we recognize that oxidation of the~$\ce{UF4}$ may also have occurred as it has been exposed to air moisture, but due to the diffusiveness of the Al/F/O peaks, this is difficult to quantify.

The elemental composition of the JYFL source (Fig.~\ref{fig:simnra_rults_JYFL}) was similarly attained by iteratively simulating multiple layers. The best match for the stainless steel substrate was obtained with a mixture of 81.5\% Fe,  18\% Cr and 0.5\% Mo. The uranium layer needed a considerable amount of oxygen in order to replicate the experimental spectrum as well as the peak in the 500~keV region. Additionally, in order to get the uranium peak width and position correct as well as the substrate, some carbon needed to be added. ``Carbon'' does not necessarily represent the element carbon but can express any light element below oxygen. Finally, the zinc layer is positioned on the surface. No layer structure, that would exactly match to all experimental spectra, was found and therefore a slight overestimation of the~$^{233}$U peak in the simulated curve is present.

The Simnra simulations estimate the~$^{233}$U layer thickness to be~$\sim\!54\cdot10^{15}$~at./cm$^2$, which is about 73\% of the calculated thickness. Variation in the thickness can be expected from the uneven coloration of sources, which can be clearly seen in the photograph of the source in Table~\ref{tab:source_properties}. The dark discoloration also supports the assumption of considerable oxidation of the uranium as indicated by the oxide peak in the RBS spectra according to which the uranium layer has 2 to 3 times the number density of oxygen than that of uranium. When exposed to room temperature air, uranium metal oxidizes rapidly so that within 3-4 days the surface of uranium has changed from a metallic coloration to a surface with a black appearance~\cite{yemel-book}. As this layer is not protective, the~$^{233}$U layer in the source has probably oxidized throughout. 

\subsection{Layer order in the JYFL source}

\begin{figure}
    \centering
    \includegraphics[width=\columnwidth]{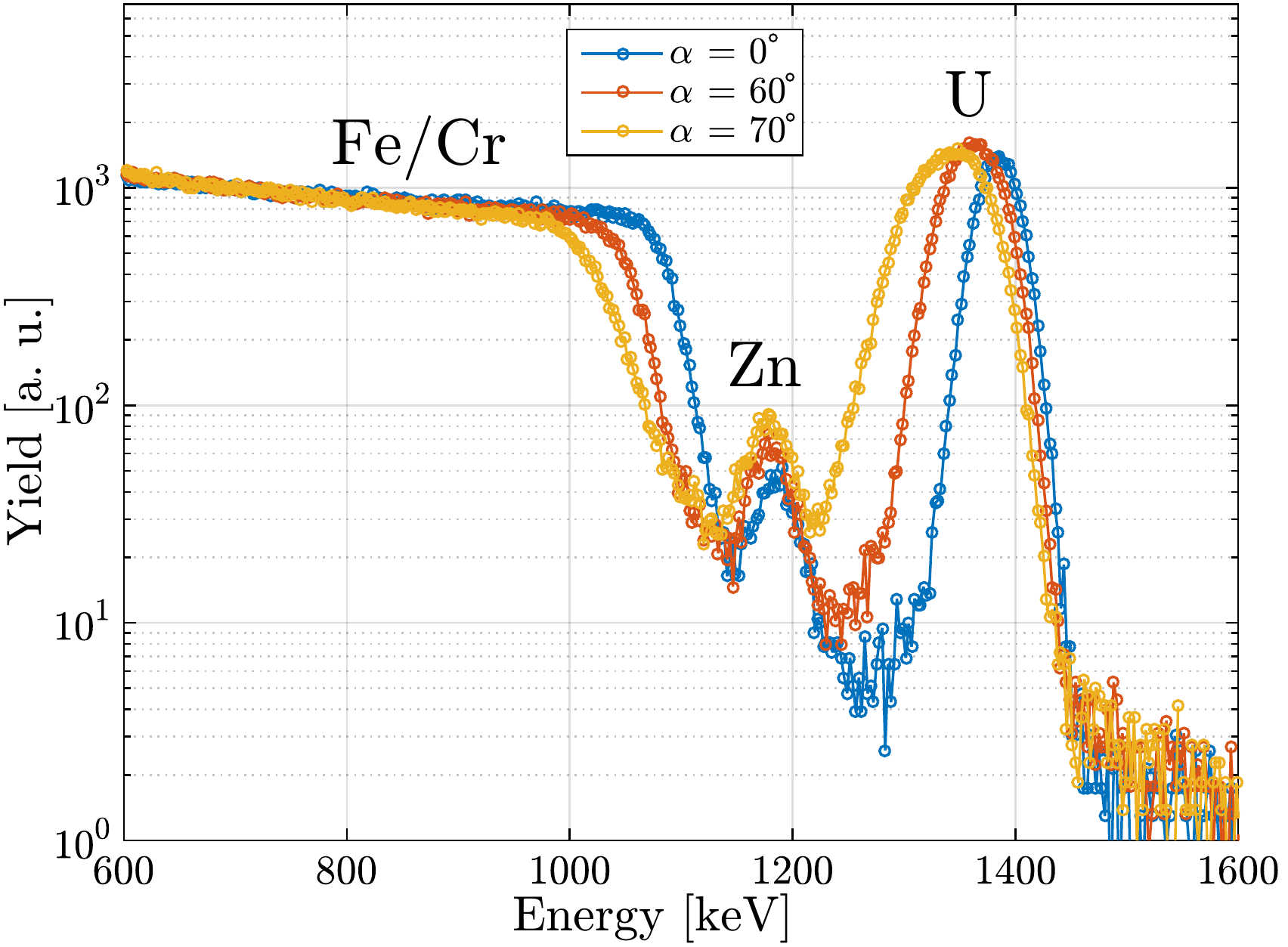}
    \caption{The experimental RBS spectrum of a JYFL source strip shown at three different sample tilt angles. The~$\alpha = 60\degree$ and~$70\degree$ spectra are normalized to the~$\alpha = 0\degree$ spectrum.}\label{fig:rbs_jyfl_tilt_spra}
\end{figure}

\begin{figure}
    \centering
    \includegraphics[width=\columnwidth]{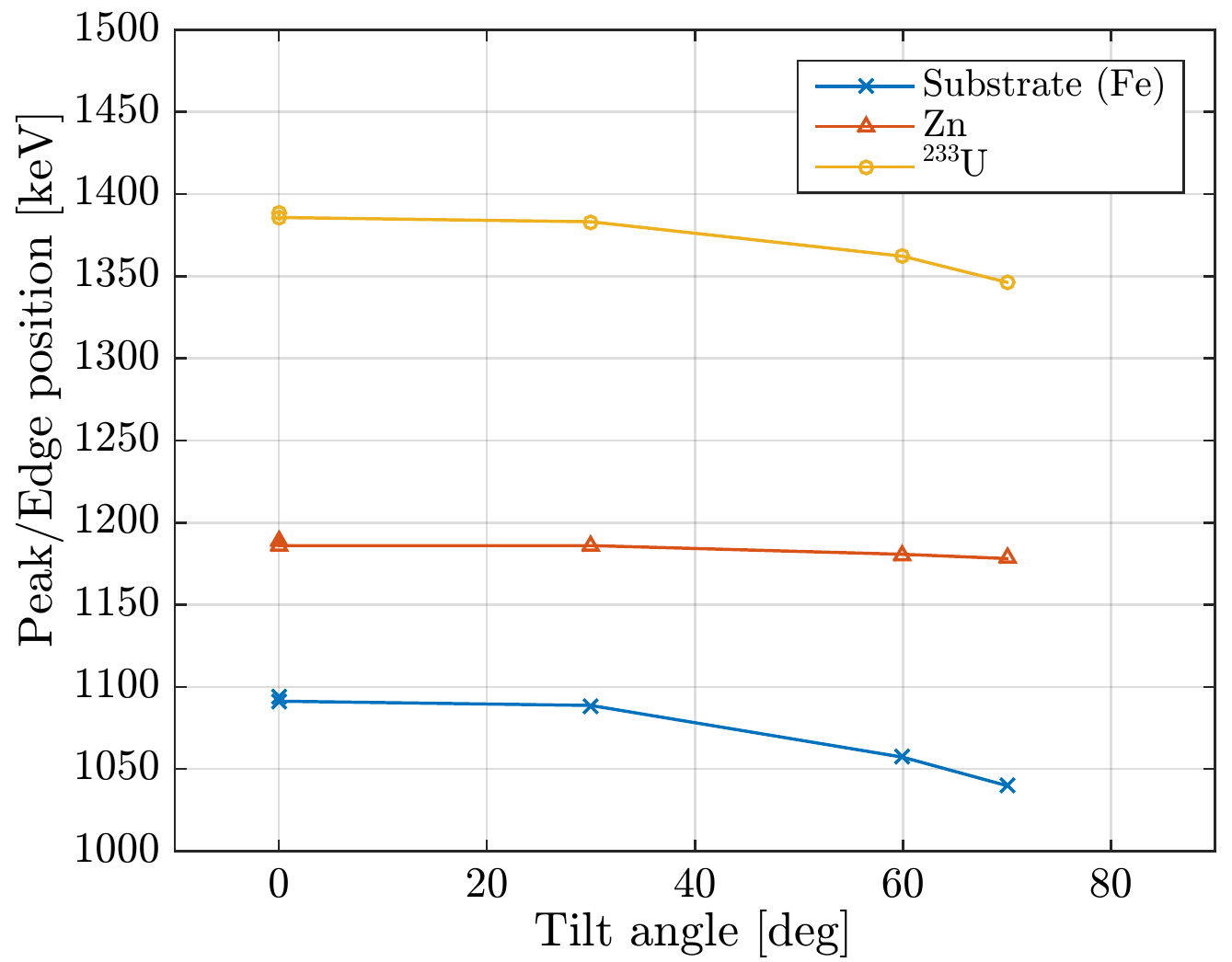}
    \caption{The peak position for the $^{233}$U, the Zn layer and the substrate edge as a function of tilt angle.}\label{fig:rbs_jyfl_peak_pos}
\end{figure}

The unexpected zinc layer was already seen in a set of earlier RBS measurements with a different source strip confirming its presence in multiple strips. In this previous set of measurements a different geometry was used,~$\beta + \alpha + \theta = 180\degree$, with the detector~$\theta = 165\degree$ and with four different tilt angles~$\alpha = 0\degree, 30\degree, 60\degree$ and $70\degree$. This allowed the arrangement of the layers and elements to be determined. In Fig.~\ref{fig:rbs_jyfl_tilt_spra} the experimental RBS spectra are shown for three different angles. As the tilt angle is increased, the peak positions move toward lower energy because the backscattered helium ions need to travel through more material in order to get to the detector which is fixed at the angle~$\theta$. By comparing the movement of the peaks and the substrate edge at different tilt angles the ordering of the layers can be inferred. Because the peak originating from the zinc layer does not move nearly as much as the~$^{233}$U or the stainless-steel substrate, as illustrated in Fig.~\ref{fig:rbs_jyfl_peak_pos}, the zinc layer has to be positioned on top of the uranium.

Finally, the composition of the back side of the JYFL source strip was measured. As expected only the stainless steel substrate was detected and no other element, including zinc, was found.

Although the simulations estimated the thickness of the zinc layer to be only 1$\cdot10^{16}$-2$\cdot10^{16}$~at./cm$^{2}$, which is capable of stopping approximately a 10~keV thorium ion, the fact that a contaminant layer was detected on top of the source is worrying. In particular because there seems to be some other ``light'' element on top, it is suspected that the low measured recoil efficiency is due to the surface contaminants. The zinc suspiciously points to the electron emitter gas cell which was used in earlier work ~\cite{Sonnenschein2012} because the components of the housing of the hot filament, used to create the high flux of electrons, are made of brass, an alloy metal containing approximately 50\% zinc.

\section{Summary and outlook}

To obtain the radiogenic composition of the~$^{233}$U sources, alpha and gamma spectroscopy measurements have been demonstrated to be a successful method. Even though~$^{229}$Th has a relatively long half-life, gamma spectroscopy of implantation foils can be used to determined the recoil efficiency if the background is controlled and long measurement times are used. The RBS technique proved to be a very useful method to probe the composition and structure of the sources and to elucidate the reasons behind the measured low recoil ion efficiencies. The analysis and methods presented in this work are not only applicable to our~$^{233}$U sources but can be applied to any such alpha recoil source or to radioactive targets used in an on-line experiment, particularly when the reaction product has a low kinetic energy.

It is clear from our work and that of L.v.d Wense and colleagues~\cite{Wense2015}, a thinner $^{233}$U source layer is advantageous for a higher recoil ion efficiency. However, using exceedingly thin sources in order to achieve high recoil efficiencies ($>$20\%) will not produce the highest recoil yield per unit area because it means that the thickness is smaller than the stopping distance of a recoil. To obtain the optimum yield, sufficient source material should be used to utilize the whole thickness of the source layer while making sure that the source material does not contain any other element or isotope other than the source isotope itself. The purity of the source material should be considered the most important factor when pursuing the highest yield sources.

\begin{figure}
  \centering
  \includegraphics[width=\columnwidth]{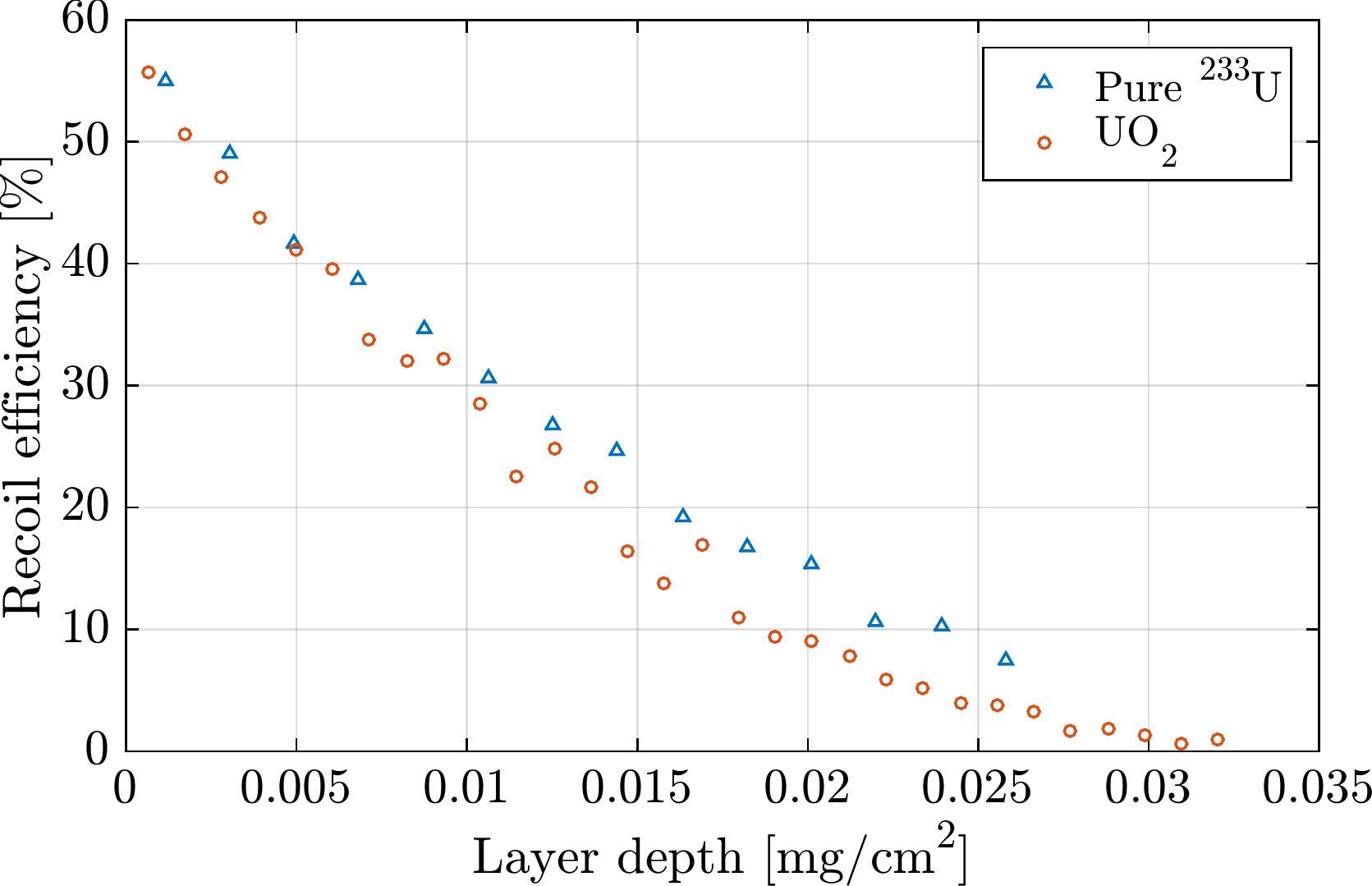}
  \caption{The simulated~$^{229}$Th recoil efficiency in pure~$^{233}$U and in~$\ce{UO2}$ as a function of depth of the point of origin for the~$^{229}$Th emission.}\label{fig:recoil_eff_vs_depth}
\end{figure}

Taking uranium dioxide ($\ce{UO2}$) as the most pure material in practice for a~$^{233}$U recoil source, an estimation for the minimum source thickness that utilizes the whole source layer for~$^{229}$Th recoils has been estimated from the TRIM simulations. Fig.~\ref{fig:recoil_eff_vs_depth} shows the recoil efficiency as a function of depth at which the~$^{229}$Th recoil ion is emitted. From the very top layer of the source efficiencies over 50\% are obtained due to the random walk-type of behavior of a~$^{229}$Th ion. In this layer the ion has a chance to escape even if emitted over 90$\degree$ away from the surface normal.~$^{229}$Th ions emitted deeper from a source have a lower probability for escaping until the probability reaches 0\% at a depth of about 0.035~$\mathrm{mg/cm^2}$ ($\sim$80$\cdot10^{15}$~$^{233}$U at./cm$^2$), which marks the layer thickness above which no more recoils are obtained even though source layer may be thicker. Using these results and the fact that the total recoil efficiency for such a $\ce{UO2}$ layer is about 16\%, when considering recoil energies above 10 keV, the largest possible~$^{229m}$Th isomer recoil emission rate in practice from a~$^{233}$U source is about 35 ions/s/cm$^2$. Taking into account a realistic gas cell extraction and purification efficiency of about 16\%~\cite{Wense2015} and the requirement of 1000~ions/s for a fluorescence-based laser spectroscopy experiment, a minimum requirement  of about 180~cm$^2$ for the surface area of a~$^{233}$U source is obtained.

As the preparation of thin $^{233}$U sources is a critical aspect to this work, it is now being supported within the nuClock collaboration~\cite{Nuclock_website} by the Institute for Atomic and Subatomic Physics of TU Wien which will provide new molecular-plated~$^{233}$U sources in the near future. There will also be future measurements in which the RBS is performed on one of the JYFL~$^{233}$U source strips that were not used in the electron emitter gas cell in order to confirm the suspected origin of the zinc or other contaminant layers. In connection with RBS, we are also planning to perform particle-induced X-ray emission (PIXE) measurements in order to be able to better identify the contaminant materials in the sources.  The~$^{233}$U sources are also planned to be studied with respect to their installment in specially designed gas cells for the characterization of extracted ion beams. With mass separation of the ion beam the content of~$^{229}$Th can be then also measured without contamination from other recoiling species. These measurements with gas cells have already been initiated using the sources described in this work and reported in~\cite{Ilkka-thesis}.

\section{Acknowledgments}

The authors would like to thank Ludwig-Maximilians-Universität for providing us with the~$^{233}$U source used in this work. The authors gratefully acknowledge funding by the EU FET-OPEN project 664732.

\section{References}
\bibliographystyle{proceedings_emis_pohjalainen}
\bibliography{proceedings_emis_pohjalainen_references}{}   

\end{document}